\documentclass[pra,showpacs,amsmath,amssymb]{revtex4}

\usepackage{graphicx}
\usepackage{dcolumn}
\usepackage{bm}

\begin{document}

\preprint{}

\title{Wavelength dependent ac-Stark shift of the $^1S_0$--$^3P_1$ transition at 657~nm in Ca}

\author{Carsten Degenhardt, Hardo Stoehr, Uwe Sterr, Fritz Riehle}

\affiliation{Physikalisch-Technische Bundesanstalt, Bundesallee 100, 38116 Braunschweig, Germany}

\author{Christian Lisdat}

\affiliation{Institut f\"ur Quantenoptik, Universit\"at Hannover, Welfengarten 1, 30167 Hannover, Germany}

\date{\today}

\begin{abstract}
We have measured the ac-Stark shift of the $4s^{2}\;^{1}S_0-4s4p\;^3P_1$ line in $^{40}$Ca for perturbing laser wavelengths between 780~nm and 1064~nm with a time domain Ramsey-Bord\'e atom interferometer. We found a zero crossing of the shift for the $m_S=0-m_P=0$ transition and $\sigma$ polarized perturbation at 800.8(22)~nm. The data was analyzed by a model deriving the energy shift from known transition wavelengths and strengths. To fit our data, we adjusted the Einstein $A$ coefficients of the $4s3d\;^3D-4s4p\;^3P$ and $4s5s\;^3S-4s4p\;^3P$ fine structure multiplets. With these we can predict vanishing ac-Stark shifts for the $^1S_0\;m=0- {^3P_1}\; m=1$ transition and $\sigma^-$ light at 983(12)~nm and at 735.5(20)~nm for the transition to the $^3P_0$ level.
\end{abstract}
\maketitle
\section{Introduction}
The realization of the SI second by the ground state hyperfine transition in Cs yields the by far best realization of all SI units. Nevertheless, some optical frequency standards based on single ion traps \cite{ytt, rafac00} or cold neutral atom clouds \cite{wil02,oat99} are about to exceed the best microwave clocks in stability and accuracy. Due to the improved capabilities of frequency transfer with fs-combs \cite{udem01,steng02} these standards can be evaluated more easily. Thus fast progress can be expected. Recently, Katori and coworkers proposed and started to explore a so called ``optical lattice clock'' \cite{ido03,kat03,taka03}. It combines the advantages of ion and neutral atom frequency standards by using large particle numbers and a confining optical potential, which allows long interrogation times and eliminates the Doppler effect and related uncertainty contributions. The wavelength and polarization of the optical trapping field are tailored such that both levels of the clock transition have the same Stark shift. Hence, the transition energy can be observed without perturbations by the trapping field. The exact parameters have to be determined experimentally with high accuracy to meet the requirements of a lattice clock, since up to now calculations do not reach the necessary precision.\\
In this article we present measurements of the Stark shift of the $^{40}$Ca $4s^2\;^1S_0-4s4p\;^3P_1$ transition to determine wavelengths leading to vanishing frequency shift of the intercombination line. Our data is analyzed by fitting appropriate oscillator strengths to match measured and calculated Stark shifts. The transition probabilities determined in this way are compared to known values, which is especially of interest in cases where only theoretical results are available.
\section{\label{theo}Theoretical description}
\begin{figure}
\centerline{\includegraphics[width=8cm]{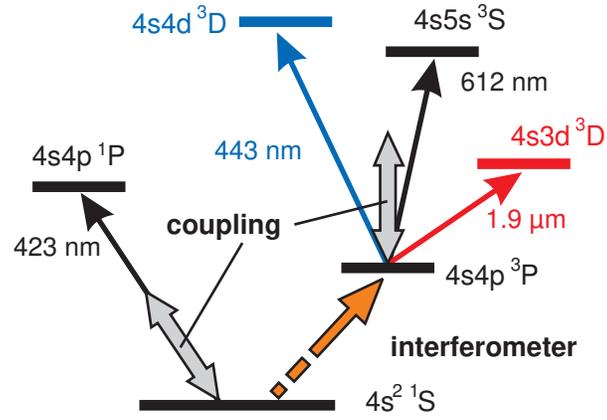}}
\caption{Simplified level scheme of the lowest levels in
$^{40}$Ca and couplings by the perturbation laser. The energies
are not to scale.} \label{fig:theo1}
\end{figure}

\begin{figure}
\centerline{\includegraphics[width=8cm]{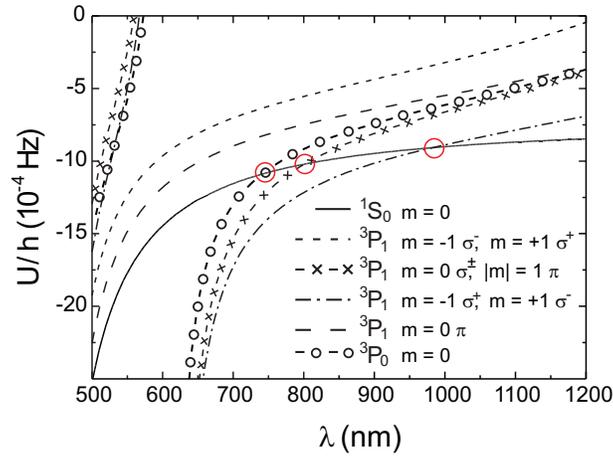}}
\caption{ac-Stark shift $U$ for a laser irradiance of 1~W/m$^2$.
Some combinations of Zeeman levels and polarizations show the
same shift. The poles at 612~nm in two curves of the $^3P_1$
manifold and the $^3P_0$ curve are due to coupling to the
$4s5s~^3S$ level. Circles denote points of vanishing frequency
shift of the intercombination line, so called ``magic
wavelengths''.} \label{fig:theo2}
\end{figure}

The energy shift $U_i(\omega,p,m_i)$ of an atomic state $i$ with energy $E_i$ and Zeeman level $m_i$, which is induced by a perturbing laser field with frequency $\nu = \omega /2\pi$, polarization $p$, and irradiance $I$ can be expressed in second-order perturbation theory as $U_i(\omega,p,m_i) = -\alpha_i(\omega,p,m_i)I/2\epsilon_0c$ with the induced polarizability $\alpha_i$. We calculated the latter by summing the contributions from all dipole transitions from the desired state $i$ to levels $k$ with the respective Einstein coefficients $A_{ki}$ (spontaneous emission rate for $E_k > E_i$), Zeeman levels $m'$, and transition frequencies $\nu_{ik}=\omega_{ik}/2\pi$ \cite{Gri00}
\begin{equation}
\alpha_i = 6\pi c^3 \epsilon_0  \sum_{k,m'}  \frac{A_{ki}(2J_k+1)} {\omega_{ik}^2(\omega_{ik}^2-\omega^2)}
\left( \begin{array}{ccc}
        J_i & 1 & J_k \\
        m_i & p & -m' \\
\end{array}\right)^2.
\label{eq:alpha}
\end{equation}
The expression in large brackets denotes a $3J$-symbol \cite{Edmonds}. It describes the selection rules and relative strengths of the transitions depending on the involved angular momenta $J$, their projections $m$ and the polarization $p$.
The transition frequencies and $A_{ki}$ coefficients were taken from the comprehensive data collection of Kurucz
\cite{kurucz}. Data from other sources, both, theoretical \cite{Mit93,Hansen,merawa01,fisch03} and experimental \cite{NIST,smi75,Yan02}
will be discussed in Sec.~\ref{res}. Additionally, the influence of the continuum and highly excited states was
approximated by using hydrogen wavefunctions \cite{Bethe} similar to the method of von Oppen
\cite{OppCont}.\\
A simplified level scheme is shown in Fig.~\ref{fig:theo1}. In Fig.~\ref{fig:theo2} the function $U_i(\omega,p,m_i)$ is depicted for different polarizations, Zeeman levels, selected levels and wavelengths. One observes that at the so called ``magic wavelength'' \cite{taka03} (for appropriate polarization of the perturbing field) the shifts of $4s^2~^1S_0$ and $4s4p~^3P_{1}$ states are the same and thus no frequency shift of the intercombination line is observed. This occurs for $\sigma$ polarized light and the $m_P=0$ Zeeman sublevel of the $^3P_1$ level around 800~nm,
and at about 1000~nm for the $m_P=+1$ ($-1$) level and $\sigma^-$ ($\sigma^+$) light. In the fermionic isotope, the transition $^3P_0- ^1S_0$ is allowed due to hyperfine level mixing. For this transition a ``magic wavelength'' exists around 735~nm. \\
The ac-Stark shift of the $^3P_{1}$ and $^3P_0$
levels is dominated in the depicted wavelength interval by the 1.9~$\mu$m, 612~nm, and 443~nm transitions shown in Fig.~\ref{fig:theo1}. Due to a vanishing value of the 3$J$-symbol in Eq.~\ref{eq:alpha} some combinations of Zeeman levels and polarizations do not couple to the $^3S_1$ level by the 612~nm transition. E.\,g.~in the case of $\sigma^-$-polarization, the $m=-1$ sublevel only couples to a $m=-2$ sublevel, which is not present in the $^3S_1$ level and thus the energy-shift $U_i(\omega,p,m_i)$ for this combination has no pole at that wavelength.\\
For the prediction of the ``magic wavelengths'', the 1.9~$\mu$m transition is of special interest, as its transition wavelength is the only one larger than the ``magic wavelengths''. Thus, only this line introduces negative contributions to $\alpha_i(\omega,p,m_i)$ in Eq.~\ref{eq:alpha}. Therefore, its transition strength has, together with the 612~nm transition, a crucial influence on the slope of the curves in Fig.~\ref{fig:theo2}. The ``magic wavelengths'', depicted by circles in Fig.~\ref{fig:theo2}, are sensitively influenced by the counterplay of both $A$-coefficients. The position of the crossings depends also on the ac-Stark shift of the $^1S_0$~level, which is dominated by the 423~nm $^1S_0 - ^1P_1$ transition. But its Einstein coefficient is well known from photoassociation spectroscopy \cite{PAgoetz} and the shift of the $^1S_0$ level is therefore well known in this wavelength region. Unfortunately, the published Einstein coefficients of the 1.9~$\mu$m line differ up to a factor of three (see Tab.~\ref{tab:osz}).\\
\section{\label{exp} Experiment}
To determine the light-induced shift of the $^1S_0 - ^3P_1$
intercombination line in $^{40}$Ca one has to compare its
transition frequency in presence of the perturbing laser field to
the frequency in the non-perturbed case. We measured the
transition frequency on a sample of laser-cooled atoms in a
magneto-optical trap (MOT) by time-domain Ramsey-Bord\'e atom
interferometry. The experimental setup as well as the use of atom
interferometry
for the realization of an optical frequency standard were described previously \cite{wil02,oat99,steng01,rie99,rie99a}.\\
In the experiment, 10$^{7}$ calcium atoms were loaded from a thermal beam into a MOT and laser cooled within 20~ms to
about 3~mK on the strong 423~nm transition from the ground state to the $^1P_1$ level. The quadrupole field and laser
beams of the MOT were then switched off, a homogeneous magnetic bias field was applied to separate
the Zeeman components of the triplet state by 3.8~MHz each, and the intercombination line was interrogated by two
counter-propagating laser pulse pairs during the expansion of the cold atomic cloud. The pulse separation within one
pair was 216.4~$\mu$s. The light was generated by a diode laser in Littman configuration, which was locked by the
Pound-Drever-Hall method to a reference resonator. A line width of below 5~Hz and frequency drift below 0.5~Hz/s were achieved. This laser served as master laser, whereas for each direction of pulse pairs, an
injection-locked slave diode was used to increase the power. The interrogation laser was stabilized to the central fringe of the
Ramsey-Bord\'e interference pattern (FWHM: 1.16~kHz) by adjusting the frequency offset between laser and reference
cavity with an acousto-optical modulator.\\
To measure the frequency shift due to the perturbing laser, we interleave two stabilization schemes with and without perturbing laser. The difference between the respective offset frequencies is the ac-Stark shift $\Delta\nu$ of the intercombination line. This method suppresses systematic
errors, e.\,g.~frequency shifts due to misalignment of the interferometer.\\
To investigate the dependence on the wavelength of the perturbation, radiation from different lasers was used. Near 780~nm, we applied a diode laser with semiconductor amplifier ($P=300$~mW at the atomic cloud). A multi-mode fiber-couple diode bar was used at 810~nm ($P=900$~mW). At 860, 890, and 922~nm, a Ti:Sapphire laser delivered around $P=250$~mW, and for 1064~nm, a Nd:YAG laser was available ($P=400$~mW). The laser power was measured with wavelength-insensitive thermal power meters, which were calibrated at 800~nm.\\
The radius $w$ ($1/e^2$ in irradiance) of these perturbing laser beams was typically in the order of 1~mm, while the
$1\sigma$-radius $R$ of the atomic cloud at the end of the trapping phase was about 0.4~mm. Under these conditions
observed shifts are in the order of 10~Hz. The size of the beam is a compromise between high irradiance to achieve large
frequency shifts and a moderate spatial inhomogeneity of the perturbation in the region of the atomic cloud. To take into account the residual inhomogeneity of the irradiance profile of the laser beam, we averaged over the beam and trap geometry. The laser beam was described as propagating in $x$-direction with constant radii $w_{y}$ and $w_{z}$ and Gaussian profile $I(y,z)$ in radial direction. The measured density of the atomic cloud was well described as three dimensional Gaussian profile $D(x,y,z)$ with $R_x \equiv R_y \neq R_z$.   The observed frequency shift $\Delta\nu$ is
\begin{eqnarray}
\Delta\nu &=& \frac{\int dx \: dy \: dz\: s(\omega,p,m_P) I(y,z) D(x,y,z)}{\int dx\: dy\: dz\: D(x,y,z)} \nonumber \\
&=& s(\omega,p,m_P)\cdot  2P /(\pi(w_y^2+4R_y^2)^{1/2}(w_z^2+4R_z^2)^{1/2}). \label{eq:Dn}
\end{eqnarray}
Here, we utilize the function
\begin{equation}
s(\omega,p,m_P) = \left[\alpha_{P}(\omega,p,m_P) - \alpha_{S}(\omega,p,0)\right]/(2\epsilon_0 h c),
\label{eq:s}
\end{equation}
which describes the frequency change of the intercombination line per unit irradiance of the perturbing
field.\\
Eq. \ref{eq:Dn} describes well the functional dependence on the geometrical quantities, as has been confirmed experimentally. However, it does not take into account the geometry of the spectroscopy and detection laser beams and the expansion of the atomic cloud. All influences were modeled by a Monte-Carlo simulation, leading to an additional correction depending on the actual conditions, but being always below ten percent.\\
The experimental setup required to have the perturbing beam propagating perpendicularly to the quantization axis given by the magnetic bias field. In this geometry, linearly polarized light with its electric field vector parallel to the quantization axis is described as $\pi$-polarization, while with 90$^\circ$ rotated polarization one realizes $\sigma$-polarization, which can be decomposed into $\sigma^{\pm}$-polarization.\\
For the $m_S=0 \rightarrow m_P=0$ transition and $\pi$-light we find an easy to interpret situation, while for $\sigma^\pm$ two field components have to be considered simultaneously. For symmetry reasons of the $3J$-symbol in Eq.~\ref{eq:alpha}, $\sigma^+$- and $\sigma^-$-light generate the same shifts on the $m=0$ levels. Hence, the sum of these shifts is the same as for pure $\sigma^+$- or $\sigma^-$-light.\\
Fluctuations of the bias field prevented the direct observation of frequency shifts in the 10~Hz range of the $m_P=\pm1$ Zeeman levels. Therefore, the spectroscopy laser was locked to the cross-over signal between the transitions to the $\pm 1$ Zeeman components. For this purpose, the bias filed was reduced to allow for overlapping Doppler profiles of both Zeeman levels. Nevertheless, the measurements showed increased frequency noise. The measured frequency shifts for $\sigma$-polarized light determined in this way can be calculated as the average of the shifts for the transitions to the $m_P=1$ with $\sigma^+$- and $\sigma^-$-light. From measurements of the $m_P=1$ cross-over with $\pi$-light no new information would be gained here, because the ac-Stark shift is the same as for $m_P=0$ with $\sigma$-light.
\section{\label{res}Results and discussion}
\begin{figure}
\centerline{\includegraphics[width=8cm]{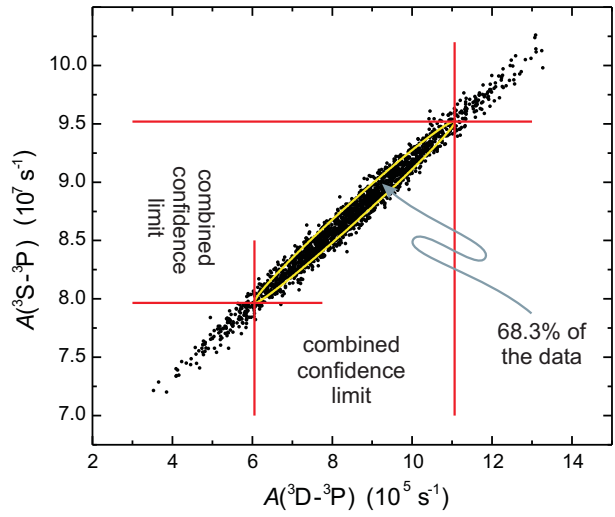}}
\caption{Einstein coefficients $A(^3D-{^3P} )$ and $A(^3S-{^3P}$)
fitted to 2000 artificial data sets generated by a Monte-Carlo
approach. Combined confidence limits for both parameters are
shown in the graph.} \label{fig:corr}
\end{figure}

\begin{figure}
\centerline{\includegraphics[width=8cm]{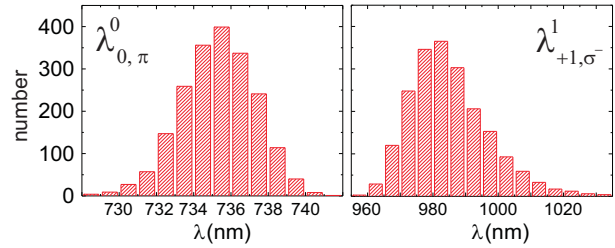}} \caption{Number
distribution for the ``magic wavelengths'' $\lambda_{0,\pi}^0$
and $\lambda_{+1,\sigma^-}^1$ as determined by the Monte-Carlo
simulation.} \label{fig:mag}
\end{figure}

The experiments described in the previous section yielded the values of $s(\lambda,p,m_P)$ depicted in Fig.~\ref{fig:meas}. The uncertainties are due to noise of the measured frequency shift, uncertainty of the measured laser power (20~mW), uncertainty of the determination of the atomic cloud size $R$ and beam radius $w$ (total contribution of 10 \%), and possibly non optimal overlap of laser beam and atomic cloud (10 \%).\\
To describe the observed ac-Stark shifts and determine the ``magic wavelength'' by a physical meaningful interpolation function (Eq.~\ref{eq:s}), the Einstein coefficients of the 1.9~$\mu$m and 612~nm transitions were fitted to the measured data (Fig.~\ref{fig:meas}). The former was chosen since it is not very well known (see Tab.~\ref{tab:osz}). It would have been sufficient to change this value to match the zero-crossing observed for the $m_P=0,~\sigma$ case, but the absolute values of $s$, especially for $m_P=0,~\pi$ would have been poorly described. Thus, $A(612$~nm) was also fitted to the data. As discussed at the end of Sec.~\ref{theo}, the interplay of $A(612$~nm) and $A(1.9~\mu$m) is crucial for the wavelength, at which the effective ac-Stark shift of the intercombination line vanishes. Other transitions are far detuned and lead to a weak dependence of the Stark shift on the wavelength in the investigated wavelength interval. Consequently, few information can be gained about other transitions.\\
Our experimental data is very well reproduced by the fit. In order to estimate the uncertainties of the so obtained ``magic wavelength'' and Einstein coefficients a Monte-Carlo simulation approach was used. Artificial data sets were generated with a statistical distribution around the values calculated from the best fit and a width of the distribution according to the experimental uncertainty. An additional random parameter with average one and standard deviation $\sigma = 0.1$ was used as common scaling factor for all data to account for systematic uncertainties like the measurement of laser power. Then, these data sets were fitted by Eq.~\ref{eq:s} and the obtained values for the $A_{ki}$'s and the ``magic wavelength'' were utilized for the error analysis.\\
Fig.~\ref{fig:corr} shows the fitted Einstein coefficients of the fine structure multiplets $A(1.9~\mu$m) and $A(612$~nm) of 2000 such iterations. A clear correlation between both parameters is visible. The combined 68.3\%-confidence limits for both parameters are also shown. The fitted $A_{ki}$'s are listed in Tab.~\ref{tab:osz} together with values from literature.  All uncertainties correspond to one standard deviation. Both $A$ coefficients agree within the uncertainties with theoretical \cite{Mit93,Hansen,merawa01,fisch03} and experimental \cite{NIST,smi75,Yan02} values from other authors. For the ``magic wavelength'' $\lambda_{0,\sigma}^1$ we found $800.8(22)$~nm. Subscripts denote $m_P$ and the polarization, the superscript indicates the angular momentum $J$ of the $^3P_J$ state. The uncertainty is one standard deviation of the distribution of values $\lambda$ calculated from the artificial Monte-Carlo data and parameter sets.\\
Furthermore, the fitted parameters were cross checked by calculating the tensor polarizability $\alpha_{\mathrm{tens}}$ of the $^3P_1$ level and the difference $\Delta \alpha_{\mathrm{scal}}$ of the scalar polarizabilities of this and the $^1S_0$ state, from the expressions given by Angel and Sanders \cite{ang68}:
\begin{eqnarray}
\alpha_{\mathrm{scal}} &=& \sum_k \frac{2 \pi \epsilon_0 c^3 (2J_k+1)}{\omega_{ik}^4 (2J_i+1)} A_{ki} \\
\alpha_{\mathrm{tens}} &=& \sqrt{\frac{40 J_i (2J_i-1)}{3(2J_i+3)(J_i+1)(2J_i+1)}}
\sum_k (-1)^{J_k-J_i}  \left\{ \begin{array}{ccc}
        1   & 1 & 2 \\
        J_i & J_i & J_k \\
\end{array}\right\} \frac{3 \pi \epsilon_0 c^3 (2J_k+1)}{\omega_{ik}^4}A_{ki} \nonumber
\end{eqnarray}
The expression in curly brackets is a $6J$-symbol, analytical expressions are tabulated in e.\,g.~\cite{Edmonds}. For both values an uncertainty analysis as described above was performed. We found $\Delta \alpha_{\mathrm{scal}} = 33 (8)$~kHz/(kV/cm)$^2$ and $\alpha_{\mathrm{tens}} = 2.9 (6)$~kHz/(kV/cm)$^2$. Within the uncertainties, they agree well with $\alpha_{\mathrm{tens}} = 2.623 (15)$~kHz/(kV/cm)$^2$ as measured by Yanagimachi {\it et al.} \cite{Yan02} and the value of $\Delta\alpha_{\mathrm{scal}} = 29.874 (87)$~kHz/(kV/cm)$^2$ one can calculate with $\alpha_{\mathrm{tens}}$ from \cite{Yan02} and the Stark shift measurements of Li and van Wijngaarden \cite{Li96}. Therefore, this check confirmed the results of our fit, but it is not suited to determine $\Delta \alpha_{\mathrm{scal}}$ or $\alpha_{\mathrm{tens}}$ more precisely.\\
If these polarizabilities with their small uncertainties are included in the fit of the parameters $A$(612~nm) and $A(1.9~\mu$m), these coefficients are almost uniquely determined by the static polarizabilities and the ac-Stark data is not well described. Therefore in principle the Einstein coefficients of all transitions should be fitted to the data, which however is not feasible, as it would lead to an underdetermined set of least-square equations. To estimate the influence of other transitions, we have included one of the Einstein coefficients of the strong transitions from the $^3P$ level to the levels $4s4d~^3D$ (443~nm), $4p^2~^3P$ (430~nm), $3d^2~^3P$ (300~nm), or $4s5d~^3D$ (363~nm) in the fitted parameters. E.g. including the transition to the $4s4d~^3D$ (443~nm) level in the fit renders a good description of $\Delta\alpha_{\mathrm{scal}}$, $\alpha_{\mathrm{tens}}$, and of our measurements. $A(443$~nm) has to be changed by only -4~\% with respect to the value given in \cite{NIST}. Similar results are obtained by fitting the other coefficients. The necessary variations are -1.5~\%, -9~\%, and -20~\%, respectively. When we included more lines in the fit, no meaningful results could be obtained because of the least-square problem starts to become underdetermined.\\
Values of $\lambda_{0,\sigma}^1$, $A$(612~nm), and $A(1.9~\mu$m) determined this way do not depend on the choice of the additional line and do not differ within the uncertainties from the above discussed results. Only the uncertainty of the Einstein coefficients is reduced. This is due to a weaker correlation between $A$(612~nm) and $A(1.9~\mu$m) (compare Fig.~\ref{fig:corr}), since $A(1.9~\mu$m) strongly influences $\alpha_{\mathrm{tens}}$. The average values of both coefficients obtained in such fits are listed in the second row of Tab.~\ref{tab:osz}.\\
We have used the fitted $A$'s to calculate the second ``magic wavelength'' of  the $^3P_1$, $\lambda_{+1,\sigma^-}^1$, and the one for the $^3P_0$ level, $\lambda_{0}^0$, to which the transition is allowed in the fermionic isotope. Both were not observed directly in the experiment. The uncertainties were determined in the same way as for $\lambda^1_{0,\sigma}$. The determined values are given in Tab.~\ref{tab:magic} (uncertainties are one standard deviation), examples of the number distributions of the error analysis are shown in Fig.~\ref{fig:mag}. The results of fits with or without including the polarizabilities in the dataset do not differ within the uncertainties.
\section{\label{impl}Implications for an optical lattice clock}
Because the light shift in the $^3P_0$ state of the fermionic
isotope $^{43}$Ca is only very weakly dependent on the
polarization of the lattice laser \cite{kat03}, it seems to be
the best candidate for an optical lattice clock. However, the low
natural abundance (0.135\%) and its half-integer nuclear spin
($I=5/2$) that leads to magnetic-field sensitive transitions
could become a problem. For the use of the $^3P_1$ state in an
optical lattice clock the transition is dependent on the
polarization of the lattice laser with respect to the magnetic
quantization field. For a dipole laser of power 1 W focussed to a
waist of 30~$\mu$m a trap depth of 35~$\mu$K is achieved with
$\sigma$-polarization at the ``magic wavelength'' at 800~nm. The
shift for the ``wrong'' $\pi$-polarization amounts to $\delta_\pi
= 200$~kHz. For a typical magnetic bias field of $B = 3~\mu$T the
magnetic field mostly determines the quantization axis, and the
shift varies as $\delta_\pi \sin^2(\phi)$, where $\phi$ is the
angular deviation from the perpendicular direction. An
experimentally achievable deviation of $\phi = 10$~mrad would
amount to a shift of 20~Hz. However, as this shift depends on the
power of the trap laser, it can be identified and corrected for,
e.\,g.~by a small change in the wavelength of the trapping laser.
We have calculated the shift for the combined electric and
magnetic fields by diagonalizing the total Hamiltonian. For the
example given above we find that at a frequency shift of 7~GHz of
the trapping laser the derivative of the light shift on the
trapping laser power vanishes. Varying the power also changes the
effective quantization axis for the combined E and B-fields. Thus
the derivative depends on the power and the light shift at this
point is not precisely canceled. But even for a relatively small
field of $3~\mathrm{\mu T}$ the quantization axis is mainly
defined by the magnetic field and the resulting frequency offset
from the unperturbed line is only 40~mHz. Similar shifts have
been found for residual ellipticity of the trapping laser beam of
$4 \times 10^{-5}$. Therefore also the $^3P_1$ state could be
employed in an optical lattice clock where the residual AC-Stark
shift contributes less than $10^{-16}$ to the uncertainty.
\section{\label{}Conclusion}
\begin{figure}
\centerline{\includegraphics[width=8cm]{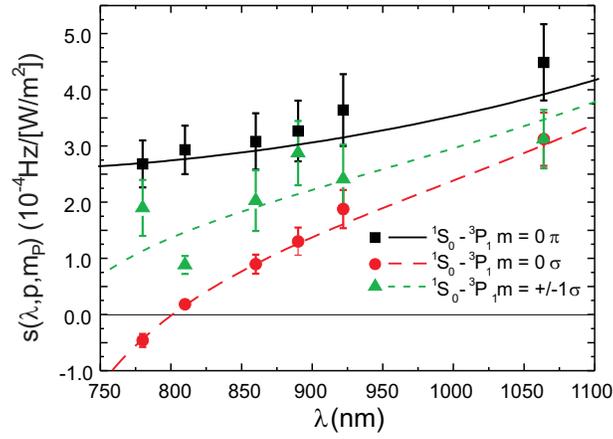}}
\caption{Measured frequency shift function $s$ measured as
described in Sec.~\ref{exp} (symbols). Lines are the best fit
according to the model introduced in Sec.~\ref{theo}. For details
about the fitting procedure see Sec.~\ref{res}.} \label{fig:meas}
\end{figure}

We have measured the ac-Stark shift of the $^{40}$Ca
intercombination line and analyzed our data by modeling the
wavelength dependence. Our data is well described after fitting
the Einstein coefficients $A$ of the transitions $4s3d\;^3D
\;-\;4s4p\;^3P$ and $4s5s\;^3S\;-\;4s4p\;^3P$ to our data
(Fig.~\ref{fig:meas}). The $A$ coefficients agree with
experimental results \cite{NIST,smi75,Yan02} and confirm recent
theoretical calculations \cite{Mit93,Hansen,merawa01,fisch03} of
the oscillator strengths (Tab.~\ref{tab:osz}). The analysis
allowed for the determination of wavelengths, at which the energy
shifts of both levels of the intercombination line are equal, the
so called ``magic wavelengths'' (Tab.~\ref{tab:magic}). We found
``magic wavelengths'' at $800.8(22)$~nm and $983(12)$~nm for the
$^3P_1$ level and $735.5(20)$~nm for the transition to the
$^3P_0$ level, which are of interest for optical lattice clocks.
We have proposed a method that is capable to reduce the influence
of the trapping field to below $10^{-16}$ even with moderate
requirements on the quality of its polarization.
\begin{acknowledgments}
The authors thank F.~Brandt for calibrating our power meters. The
loan of the lasers necessary for the experiments by the
Laserzentrum Hannover e.V., Germany and the Toptica Photonics AG
is gratefully acknowledged. This work was supported by the
Deutsche Forschungsgemeinschaft (DFG).
\end{acknowledgments}
\begin{table}
\caption{\label{tab:osz} Determined Einstein $A$ coefficients (first two lines) and values from literature. The last column indicates theoretical or experimental origin of data.}
\begin{ruledtabular}
\begin{tabular}{cccc}
reference & $A(4s3d\;^3D \;-\;4s4p\;^3P)$ & $A(4s5s\;^3S\;-\;4s4p\;^3P)$ & origin \\
\hline
this work \footnotemark[1] & 8.6(25)$\cdot10^5$~s$^{-1}$ & 8.7(8)$\cdot10^7$~s$^{-1}$ & exp.\\
this work \footnotemark[2] & 7.8(4)$\cdot10^5$~s$^{-1}$  & 8.5(4)$\cdot10^7$~s$^{-1}$ & exp.\\
\cite{kurucz}   & 3.7$\cdot10^5$~s$^{-1}$     & 6.6$\cdot10^7$~s$^{-1}$ & theo./exp.\\
\cite{NIST}     &                             & 8.6$\cdot10^7$~s$^{-1}$ & exp.\\
\cite{Mit93}    & 9.2$\cdot10^5$~s$^{-1}$     & 8.3$\cdot10^7$~s$^{-1}$ & theo.\\
\cite{Hansen}   & 9.1$\cdot10^5$~s$^{-1}$     & 8.1$\cdot10^7$~s$^{-1}$ & theo.\\
\cite{merawa01} & 3$\cdot10^5$~s$^{-1}$       & 7.8$\cdot10^7$~s$^{-1}$ & theo. \\
\cite{fisch03}  & 8.4(25)$\cdot10^5$~s$^{-1}$ & 8.6(9)$\cdot10^7$~s$^{-1}$ & theo.\\
\cite{smi75}    &                             & 8.6(13)$\cdot10^7$~s$^{-1}$ & exp.\\
\cite{Yan02}    & 7.06 $\cdot10^5$~s$^{-1}$   & & exp.\\
\end{tabular}
\end{ruledtabular}
\footnotetext[1]{Fitted without static polarizabilities $\Delta \alpha_{\mathrm{scal}}$, $\alpha_{\mathrm{tens}}$.}
\footnotetext[2]{Fitted including static polarizabilities and one additional line (see text).}
\end{table}
\begin{table}
\caption{\label{tab:magic} ``Magic wavelengths'' determined in this work. $m_P$ denotes the upper state's Zeeman level, the polarization given belongs to the perturbing laser.}
\begin{ruledtabular}
\begin{tabular}{cccc}
transition & $m_P$ & polarization & value \\
\hline
$4s^2~^1S_0-~4s4p~^3P_1$ & 0 & $\sigma$ & 800.8(22)~nm \\
$4s^2~^1S_0-~4s4p~^3P_1$ & +1 ($-1$) & $\sigma^-$ ($\sigma^+$) & 983(12)~nm \\
$4s^2~^1S_0-~4s4p~^3P_0$ & 0 & arbitrary & 735.5(20)~nm \\
\end{tabular}
\end{ruledtabular}
\end{table}

\enddocument